\documentclass[11pt,a4paper]{article}
\pdfoutput=1
\usepackage{jcappub}
\usepackage{latexsym,amsmath,amssymb}
\usepackage{slashed}
\usepackage{bm}
\usepackage{mathrsfs}

\newcommand{\nn}{\nonumber}
\newcommand{\bea}{\begin{eqnarray}}
\newcommand{\ea}{\end{eqnarray}}
\newcommand{\rar}{\rightarrow}
\newcommand{\Rar}{\Rightarrow}

\def\dd{\text{d}}
\def\d{\partial}
\def\+{\dagger}

\newcommand{\cph}{\varphi}
\newcommand{\cth}{\vartheta}
\newcommand{\cep}{\varepsilon}

\newcommand{\cA}{{\cal A}}
\newcommand{\cB}{{\cal B}}

\newcommand{\cH}{{\cal H}}

\newcommand{\cP}{{\cal P}}

\newcommand{\cS}{{\cal S}}
\newcommand{\cW}{{\cal W}}

\newcommand{\Mpc}{\text{Mpc}}
\newcommand{\GeV}{\text{GeV}}

\newcommand{\nGauss}{\text{nGauss}}

\newcommand{\vk}{{\mathbf k}}
\newcommand{\vp}{{\mathbf p}}
\newcommand{\vq}{{\mathbf q}}

\newcommand{\vx}{{\mathbf x}}

\newcommand{\vz}{{\mathbf z}}

\def\la{\langle}
\def\ra{\rangle}

\title{Pseudoscalar N-flation and axial coupling revisited}

\author{Federico~R.~Urban}

\affiliation{Service de Physique Th\'eorique, Universit\'e Libre de Bruxelles, CP225, Boulevard du Triomphe, B-1050 Brussels, Belgium}

\emailAdd{furban@ulb.ac.be}

\date{\today}

\abstract{We revisit the dynamics of the axial coupling between many N-flatons and an Abelian gauge field, with special attention to its statistically anisotropic signal.  The anisotropic power spectrum of curvature perturbations associated to the large wavelength modes of the gauge vector field is generally undetectable, since the anisotropy is confined to small scales.  If the gauge field is the electromagnetic field, provided that the number of fields participating in the exponential expansion is large, it could be possible to generate sizable large scale magnetic fields.  However, its spectrum is blue, and appreciable power on large scales implies an overly strong field on smaller scales, incompatibly with observations.  Furthermore, the anisotropy is also markedly enhanced, and might be at odds with the isotropic observed sky.  These aspects further demand that the scale of inflation is kept to a minimum.}

\arxivnumber{}

\begin{document}
\maketitle

\section{Foreword, with results}\label{s:axe}

Statistical isotropy is one of the basic assumptions of the Standard model of Cosmology, beautifully verified by Cosmic Microwave Background observations~\cite{Ade:2013zuv}.  The recipe for generating the inhomogeneities which populate the early Universe, and which evolve in the observed gravitational structures, comes from Inflation, an epoch of accelerated expansion (see for instance~\cite{Mukhanov:2005sc}).  Typically, inflation is a very isotropic affair, owing to the scalar nature of the background field driving the expansion.  However, it is possible to envision the presence of more fields; in fact, it is desirable to include them for the de Sitter expansion has to come to an end, and some kind of interactions are necessary to repopulate the vacuum.  One is led to consider scalar as well as higher spin fields.

A vector spin one field which is dynamic during inflation can source statistically anisotropic classical curvature perturbations, see~\cite{Yokoyama:2008xw,Bartolo:2012sd,Lyth:2013sha} (a fully anisotropic study of this effect is available in~\cite{Watanabe:2010fh,Abolhasani:2013zya}).  Interesting in its own right nonetheless, after the results of Planck~\cite{Ade:2013nlj}, hinting indeed at a statistically anisotropic spectrum, such possibility garnered extra credit.

In this work we calculate the statistical anisotropy generated by the axial coupling of a (collection of) pseudoscalar inflaton to a $U(1)$ gauge field~\cite{Turner:1990uz,Pospelov:2008gg,Barnaby:2010vf,Barnaby:2011vw}.  Such coupling has been studied in connection to magnetogenesis models (when the $U(1)$ is identified with the electromagnetic field)~\cite{Turner:1987bw,Garretson:1992vt,Anber:2006xt,Durrer:2010mq}.  The vector field in this study will always be subdominant, and is assumed to not alter the isotropy of the background solution to a significant extent; we thence employ the standard Friedmann-Lema\^itre-Robertson-Walker metric.

\subsection{The anisotropy}

A statistical anisotropy is unavoidably generated in this setup.  This because at each horizon crossing any given curvature perturbation mode will encounter a (subdominant) classical background of infrared vectors which left before it.  These modes stack up as once they abandon the horizon they classicalise, and there is then a preferred direction for each realisation of inflation, which belongs to a given (nearly Gaussian) distribution.  The superhorizon evolution of each quantum mode then takes place in an anisotropic (albeit slightly so) background; this will reflect onto the final spectrum of the gauge field perturbations.

The final curvature perturbations, being themselves directly sourced by the axial coupling between the spin zero and one fields, once inflation is over, will have a two-point function (power spectrum) which also explicitly contains a directional dependence:
\[
  \la \zeta_\vp(\eta) \zeta_\vq(\eta) \ra = 2\pi^2 \cP(p) \left[ 1 + \delta\cP(\vp) \right] \frac{\delta^3(\vp+\vq)}{p^3} \, .
\]
We compute this two-point function for a simple model where the coupling is of the form
\bea\label{axe}
  {\cal L_\text{int}} = \frac14 f(\cph) F \tilde F \, ,
\ea
where the pseudoscalar function $f(\cph)$ depends on the background inflaton $\cph$ --- the simplest case prescribes $f(\cph) \sim \cph$ ---, $F$ is the gauge field two-form strength, and $\tilde F$ its dual\footnote{Notice that in fact the field(s) could as well be scalar(s); in order to preserve parity --- if this is what is wanted --- the interaction term needs to be a scalar overall, so the structure should be engineered accordingly.}.  If the coupling function is approximately linear in conformal time $\eta$ such that $\dd f / \dd \ln\eta = \text{const}$, then nearly everything can be done analytically; we will consider this possibility hereafter.

Before digging the details of the calculation we collect our main findings below.

\subsection{The correlators}

The gauge field fluctuations are boosted exponentially through the coupling with the background; however, this exponential amplification is not very efficient, mostly because the coupling itself is active only just around horizon crossing, and it is not a time-exponential enhancement (see for example~\cite{Urban:2013aka} for an example of the latter possibility).  As always in this kind of chiral couplings, of the two physical helicities of a massless $U(1)$ field, only one is amplified, the second one being instead suppressed.  We then discard the decaying solution.

We obtain, for the anisotropic two-point and three-point functions of the curvature perturbation $\zeta$:
\bea
  \la \zeta_\vp \zeta_\vq \ra &\sim& \frac{9\pi^5}{2\xi^4} \left(\frac{\rho}{\epsilon\bar{\rho}}\right)^2 \frac{p}{\cH^4} \frac{\ln^2(2\xi p/\cH)}{\ln^2(2\xi)} \sin^2\cth \, \delta^3(\vp+\vq)\label{twozEND} \, , \\
  \la \zeta_\vp \zeta_\vq \zeta_\vk \ra &\sim& \frac{\pi}{3\times2^{11} \xi^8} \left(\frac{\rho}{\epsilon\bar{\rho}}\right)^3 \frac{p q}{\cH^{8}} \frac{\ln^2(2 \xi p/\cH) \ln^2(2 \xi q/\cH)}{\ln^4(2\xi)} \mathfrak{A} \, \delta^3(\vp+\vq+\vk) + 2 \, \text{perm} \label{threezEND} \, ,
\ea
where for $\rho$ and $\bar{\rho}$ we mean the energy densities associated with the gauge field and the background N-flatons, respectively; $\epsilon$ is the first slow-roll parameter $\cH^2-\cH' \equiv \epsilon\cH^2$, where $\cH \equiv a'/a$ is the comoving Hubble parameter, $a(\eta)$ is the scale factor, and priming stands for conformal time $\eta$ derivative.  The $\xi$ controls the speed of the enhancement for the vector $U(1)$ field.  The angular dependence (thence, the anisotropy) is encoded in the angles $(\cth,\cth_p,\cth_q,\cth')$ through the $\sin\cth$ and $\mathfrak{A}(\cth_p,\cth_q,\cth')$ functions\footnote{As defined in the main text, $\cth_{(p,q,k)}$ are the angles between the vector $\bar{A}$ and $(\vp,\vq,\vk)$ and $\cos\cth' = \hat{p}\!\cdot\!\hat{q}$; $\cth=\cth_p$.}.

The isotropy of the background roughly demands that $\rho \ll \bar{\rho}$, but a stronger constraint comes from requiring that the slow-roll parameter $\epsilon$ stay small: one can verify that this translates approximately into $\rho \ll \epsilon\bar{\rho}$.  The first coefficient in both expressions is hence small.

The steep spectral dependences for both the spectrum and bispectrum make the large scales completely negligible: there is next to no anisotropy in the infrared.  In particular, for the spectrum one finds a slope close to $n_s \sim 4$ --- with a logarithmic modulation.  The reason for this behaviour is easily understood since, as we already mentioned, the coupling is only active for a short time during horizon crossing; in the absence of a strong $p$-dependence of the coupling function itself, it is clear that once outside the horizon the modes keep decaying adiabatically.  The less the time they spend in the superhorizon regime, the more the effects of the interaction are seen.  This is quite in contrast with the case of a scalar $I^2(\cph) F^2$ coupling, where in general the amplification of a given mode proceeds throughout the entire superhorizon regime, until the end of inflation: in that case large scales can dominate the power spectrum.

The logarithmic enhancements are the direct effect of the interaction term.  As we will see, in the case of a constant $\xi \equiv \eta f'(\eta)/2$, the Fourier modes of the gauge field are boosted logarithmically with time (albeit there is an overall exponential amplification with $\xi$ itself).  These $\ln$ factors appearing in~(\ref{twozEND}) and~(\ref{threezEND}) are the manifestation of this behaviour, and tend to slightly favour large wavelengths; their spectral enhancement is nonetheless small compared to the adiabatic suppression.  Notice that in general we require $\xi\gtrsim1$ or the deviations from vacuum would be even smaller.

Summarising, the curvature power spectrum does receive an isotropic contribution coming from the first order perturbed inflaton itself, and an anisotropic piece owing to the vector field; the latter is suppressed at large scales, thus preventing this anisotropy from being in the detectable range.  A strong anisotropy could develop at very small momenta, but for typical values of $H$ this will necessarily lie beyond the scales probed by observations.  There is of course the option of having a much lower energy scale of inflation: then a strongly $p$-dependent anisotropic signal develops quite quickly.  Jointly, applications to magnetogenesis also may become feasible.  This scenario is more demanding in terms of engineering (it necessitates other fields, such as curvatons), but it is a workable setup rich with signatures and applications.

We will further discuss these possibilities and all our results in the concluding section~(\ref{s:disc}).  Before that, we outline the steps of the calculation leading to the expressions~(\ref{twozEND}) and~(\ref{threezEND}) in Sec.~(\ref{s:curve}).

\section{Computing the curvature perturbation}\label{s:curve}

The system is comprised of a pseudoscalar field $\cph$ and a $U(1)$ vector potential $A^\mu$.  The field strength of the latter is as usual defined as $F_{\mu\nu}=\d_\mu A_\nu - \d_\nu A_\mu$.  The dual tensor is given by $\tilde F_{\mu\nu} \equiv \epsilon_{\mu\nu\rho\sigma} F^{\rho\sigma} /2$ with $\epsilon_{0123} = \sqrt{-g}$, $g_{\mu\nu}$ being the metric.  The coupling function $f(\cph)$ is most naturally identified with the field itself $f(\cph) \equiv \lambda \cph$ with $\lambda$ a dimensionless coupling constant; more general functions are also possible.

With a single field inflaton, perturbativity of the Lagrangean demands roughly that $\lambda\lesssim1$, and the departures from vacuum are very marginal~\cite{Durrer:2010mq}.  In the case of a multitude of pseudoscalars participating in inflation however, $\lambda$ can be large.  A large number of axion-like fields in general will follow some complicated dynamic evolution; in some cases (identical coupling for all) it is possible to assimilate the multitude to a single field with large coupling constant~\cite{Dimopoulos:2005ac}: in our setup this serves mostly for the purpose of justifying such large coupling, whose phenomenology we analyse below.

\subsection{The gauge modes}

Working in the Lorenz-Coulomb gauge for which $\d_i A^i = 0 = A_0$, it is easiest to write down the equations of motion for the coupled system in terms of the fields' Fourier components as
\bea\label{eomA}
  \cA_h'' + \left( k^2 + h k f' \right) \cA_h = 0 \, ;
\ea
we work in the helicity basis for which $\cep_h \equiv (\cep_1 \pm i \cep_2) / \sqrt2$, $h \equiv \pm$, and the Fourier transform is
\bea
  A_i(\vx,\eta) \equiv \bar{A}_i(\eta) + \sum_h \int \frac{\dd^3k}{(2\pi)^3} \cep^i_h(\vk) e^{i\vk\cdot\vx} \left( a_h(\vk) \cA_h(k,\eta) + a_h^\+(-\vk) \cA_h^*(k,\eta) \right) \, . \label{cA}
\ea
The $\cep^i_h$ are the components of the two physical polarisation vectors.

Once the coupling $f(\eta)$ is fixed, the solutions for the vector potential are known.  In the particular case of a constant $\eta f'(\eta)$ we find
\bea
  \cA_h(k,\eta) &=& \frac{1}{\sqrt{2k}} \left[ G_0(h\xi,x) + i F_0(h\xi,x) \right] \, , \label{cAsol}
\ea
where the integration constants have been determined by demanding that the Bunch-Davies vacuum be realised in the infinitely remote past $x \equiv k|\eta| \rar \infty$ --- there the mode functions reduce to plane waves $\exp(\pm i k \eta)/\sqrt{2k}$.  The $F_0$ and $G_0$ are regular and irregular Coulomb wave functions, respectively.  Only the positive helicity modes are amplified, so we can dismiss the others for what follows: $\cA \equiv \cA_+$.  Albeit in principle possible, it is not feasible to work directly with these exact expressions; only once the mode approaches and leaves the horizon does the coupling becomes effective: for large scales we have $x\ll1$, which allows writing
\bea
  G_0(\xi,x) &\simeq& 2 \cW(\xi) \sqrt{2 \xi x} K_1(2\sqrt{2 \xi x}) \nn\\
  &\simeq& \cW(\xi) \left[ 1 + 2 \xi x \left( \ln(2 \xi x) - 1 + 2\gamma \right) + \ldots \right] \, , \label{g0}\\
  &&\nn\\
  F_0(\xi,x) &\simeq& \frac{1}{\cW(\xi)} \frac{x}{\sqrt{2 \xi x}} \, I_1(2\sqrt{2 \xi x}) \nn\\
  &\simeq& \frac{1}{\cW(\xi)} \left[ x + \xi x^2 + \ldots \right] \, , \label{f0}
\ea
with
\bea\label{cW}
  \cW^2(\xi) \equiv \frac{e^{\pi\xi} \sinh\pi\xi}{\pi\xi} \stackrel{\xi\gtrsim1}{\longrightarrow} \frac12 \frac{e^{2\pi\xi}}{\pi\xi} \, ,
\ea
and where $K_1$ and $I_1$ are Bessel functions; $\gamma \approx 0.58$ is the Euler-Mascheroni constant.  Useful expressions are also the conformal time derivatives $\d_\eta = -k\d_x$:
\bea
  \d_x G_0(\xi,x) &\simeq& -4\xi\cW(\xi) K_0(2\sqrt{2 \xi x}) \nn\\
  &\simeq& 2\xi\cW(\xi) \left[ \ln(2 \xi x) + 2\gamma + \ldots \right] \, , \label{dg0}\\
  &&\nn\\
  \d_x F_0(\xi,x) &\simeq& \frac{1}{\cW(\xi)} \, {}_0F_1(1,2 \xi x) \simeq \frac{1}{\cW(\xi)} \, I_0(2\sqrt{2 \xi x}) \nn\\
  &\simeq& \frac{1}{\cW(\xi)} \left[ 1 + 2 \xi x + \ldots \right] \, . \label{df0}
\ea
Again the $K_0$ and $I_0$ are Bessel functions, and the ${}_0F_1$ is a Hypergeometric function which simplifies to the Bessel $I_0$ in this limit.  Notice that the $G_0$ and $\d_x G_0$ terms are typically the dominant ones, so we will only consider these in the remainder of the paper\footnote{In the series of papers~\cite{Anber:2006xt,Anber:2009ua,Anber:2012du} this kind of mechanism was analysed in the case of a large number of pseudoscalars, which then allows for a large yet perturbative coupling $\xi$.  However, in~\cite{Anber:2006xt} the wrong limit of large $x\xi$ was taken, which results in different expressions for the gauge field spectrum, energy density, and its contribution to the curvature perturbation.  With the correct expansions the spectrum turns out to be strongly tilted (and anisotropic), see below.  This is not the case in~\cite{Anber:2009ua,Anber:2012du}, where their results apply within the window $1/(8\xi)\ll x \ll 2 \xi$, which always exists for $\xi\gg1/4$; the anisotropy in this case will still be present, but the spectrum is much more difficult to compute --- we discuss this briefly below.}.

The energy density stored by the $U(1)$ field can be computed as:
\[
  \rho_\cA = \frac{1}{4\pi^2 a^4} \int \frac{\dd k}{k} k^3 \left[ |\cA'|^2 + k^2 |\cA|^2 \right] \, .
\]
The momentum integration runs for those modes for which $x \lesssim 2\xi$ for $\xi \gtrsim 1$, so that the total energy density is approximated by
\bea\label{energy}
  \rho_\cA \simeq \frac{\xi^2 \cW^2 H^4}{8\pi^2} x^4 \ln^2(2 \xi x) \rar \frac{8 \xi^6 \cW^2 H^4}{\pi^2} \ln^2(2 \xi) \, .
\ea

\subsection{An equation for the curvature perturbation}

The background dynamics is governed by the Euler-Lagrange equation for the homogeneous part of $\cph(\vx,\eta) \equiv \bar{\cph}(\eta) + \delta\cph(\vx,\eta)$
\bea
  \bar{\cph}'' + 2\cH\bar{\cph}' + a^2 V_\cph = 0 \, ,\label{eom}
\ea
together with the background Einstein field equations $G_{\mu\nu} = T_{\mu\nu}$ (in units where the gravitational coupling constant is 1)
\bea
  \cH^2 = \frac13 \left( \frac12 \bar{\cph}'^2 + a^2 V_\cph \right) \, , \quad \cH^2 - \cH' = \frac12 \bar{\cph}'^2 \, .\label{bk}
\ea
Working for instance in the longitudinal gauge
\bea\label{metric}
  \dd s^2 = a^2 \left[ (1+2\Phi) \dd \eta^2 - (1-2\Psi) \dd \vx^2 \right] \, ,
\ea
the scalar curvature perturbation is defined as
\bea\label{zetadef}
  \zeta \equiv \Psi + \frac{\cH^2}{\cH^2-\cH'} \left( \frac{\Psi'}{\cH} + \Phi \right) \, .
\ea
The choice of the gauge in our case is dictated by the relative ease with which we can obtain a closed equation for the curvature perturbation $\zeta$, without approximations.  More common gauges (flat gauge, uniform density gauge) of course lead to identical results.  The governing differential equation for the evolution of $\zeta$ reads
\bea\label{zetadyn}
  \zeta'' + 2\cH \zeta' + k^2 \zeta = -\frac{\cH}{\epsilon\bar{\rho}} \left\{ \Pi' + \rho' + 6 \cH \rho \right\} \, ,
\ea
where we have used that $P = \rho/3$ for a massless field, and the momentum conservation equation $Q'/\cH = 2\Pi/3 - P - (3+\epsilon) Q$ with $Q \equiv {\rm i} \cH k^j q_j /k^2 a$ has been used; finally, $\Pi(\vk)\equiv -3 \hat k^i\hat k_j \Pi^j_i(\vk)/2$.  For the derivation of~(\ref{zetadyn}) see for instance~\cite{Bonvin:2011dt,Urban:2012ib}.  All unbarred hydrodynamic quantities refer to the gauge field in the perfect fluid decomposition with respect to the fluid velocity $u^\mu =(1-\Phi,{\bf 0})/a$
\[
  T^{\mu\nu}_{U(1)}=(\rho +P) u^\mu u^\nu - P g^{\mu\nu} + u^\mu q^\nu + u^\nu q^\mu + \Pi^{\mu\nu} \, .
\]
Barred quantities refer instead to the background inflaton.  Notice that we are disregarding second order perturbations in the metric (and higher orders in the gauge field) for they are subdominant in the present setup\footnote{There is no agreement in the literature concerning the counting of powers in the inflaton plus vector gauge field.  In particular, the latter is sometimes considered as a first order perturbation itself --- pertaining to the quantum field theoretical approach ---, or a half-order perturbation, with the understanding that in the stress tensor only quantities quadratic in the fields are allowed --- a homogeneous background is strictly speaking incompatible with the isotropic background metric.  As far as the final results are consistent with the scheme employed (i.e., higher order quantities are indeed smaller than lower order ones) the two approaches are completely equivalent.  In our case, the most relevant additional contribution to the curvature perturbations comes from the first terms in the perturbative expansion which contain the gauge field source, which we thence call first order.}.

Eq.~(\ref{zetadyn}) can be solved using the Wronskian method --- identical to the Green's function method (the Wronskian is nothing other than the Green's function itself).  In terms of ``time''\footnote{Note: this is not the configuration space coordinate $\vx$ denoted in bold.} $x\equiv-k\eta$ and using $\cH = -1/\eta$:
\bea\label{zetax}
  \zeta_{,xx}(x) - \frac{2}{x} \zeta_{,x}(x) + \zeta(x) = \cS(x) \, ,
\ea
where the source term is
\bea\label{source}
  \cS(x) = \frac{1}{\epsilon\bar{\rho}} \left[ x\Pi_{,x} + x\rho_{,x} - 6\rho \right] \, .
\ea
The solution is given as
\bea
  \zeta_\vk(x) &=& \int_{x_h}^x \dd y G(x,y) \cS_\vk(y) \, , \label{solzeta}
\ea
where
\bea
  G(x,x') &\equiv& \frac{\pi}{2} x' \left(\frac{x}{x'}\right)^{3/2} \! \left[ J_{3/2}(x')J_{-3/2}(x)-J_{3/2}(x)J_{-3/2}(x') \right] \, . \label{green}
\ea
The $J_n(x)$ are Bessel functions; since we are interested in superhorizon modes, and since small $x$ dominate the solutions for the gauge Fourier modes, the Green's function can be Taylor expanded for small $(x,x') \ll 1 \approx x_h$ to $-x'/3$.  The two-point function is obtained through
\bea\label{twoz}
  \la \zeta_\vp(x) \zeta_\vq(x) \ra &=& \int \dd x_1 \dd x_2 \, G(x,x_1) G(x,x_2) \la \cS_\vp(x_1) \cS_\vq(x_2) \ra \nn \\
  &\approx& \frac19 \int \frac{\dd x_1}{x_1} \frac{\dd x_2}{x_2} \la x_1^2 \cS_\vp(x_1) x_2^2 \cS_\vq(x_2) \ra \, .
\ea
Similarly, for the bispectrum one writes
\bea\label{threez}
  \la \zeta_\vk(x) \zeta_\vp(x) \zeta_\vq(x) \ra &=& \int \dd x_1 \dd x_2 \dd x_3 \, G(x,x_1) G(x,x_2) G(x,x_3) \la \cS_\vp(x_1) \cS_\vp(x_2) \cS_\vp(x_3) \ra \nn \\
  &\approx& -\frac{1}{27} \int \frac{\dd x_1}{x_1} \frac{\dd x_2}{x_2} \frac{\dd x_3}{x_3} \la x_1^2 \cS_\vk(x_1) x_2^2 \cS_\vp(x_2) x_3^2 \cS_\vq(x_3) \ra \, .
\ea
Any n-point function follows in a similar fashion.

\subsection{Integrals of the source}

In the source terms $\rho$ and $\Pi$ there will appear in general complicated expressions bilinear in $(\bar{A}',k\bar{A})$ and $(\cA',k\cA)$.  Inspection of the explicit solutions~(\ref{cAsol}), with the help of the Taylor expansions~(\ref{g0}), (\ref{f0}), (\ref{dg0}), and~(\ref{df0}), shows that the conformal time derivative terms dominate; it then makes sense to retain only these pieces.  Schematically:
\bea
  \rho_\vk(x) &\subset& \frac{1}{2a^4} A_i'(\vx,\eta) A_i'(\vx,\eta) \label{rhox} \, , \\
  \Pi_\vk(x) &\subset& \frac{1}{2a^4} \left[ A_i'(\vx,\eta) A_i'(\vx,\eta) + 3\hat{k}_i\hat{k}_j A_i'(\vx,\eta) A_j'(\vx,\eta) \right] \label{pix} \, ;
\ea
once we substitute the Fourier expansion~(\ref{cA}) we find
\bea
  \rho_\vk(x) &\subset& \frac{1}{a^4} \bar{A}_i'(x) \sum_h \cep^i_h(\vk) \left[ a_h(\vk) \cA_h'(k,x) + a^\+_h(\vk) {\cA_h^*}'(k,x) \right] \label{pi} \\
  &-& \frac{1}{2a^4} \sum_{hh'} \int\frac{\dd^3 k'}{(2\pi)^3} \cep_h^i(\vk') \cep_{h'}^i(\vk-\vk') a_h(\vk') a_{h'}(\vk-\vk') \cA_h'(k',x) \cA_{h'}'(|\vk-\vk'|,x) + 3\,\text{h.c.} \nn
\ea

This operator possesses the anisotropic spectrum and bispectrum
\bea
  \la \rho_\vp(x_p) \rho_\vq(x_q) \ra &=& \frac{(2\pi)^3\delta^3(\vp+\vq)}{a_{x_p}^4 a_{x_q}^4 } \left[\bar{A}'(x_p)\!\cdot\!\bar{A}'(x_q) - \hat{p}\!\cdot\!\bar{A}'(x_p) \, \hat{q}\!\cdot\!\bar{A}'(x_q)\right] \cA_p'(x_p) \cA_q'(x_q) \nn\\
  &=& 2p\xi^2\cW^2 (2\pi)^3 \delta^3(\vp+\vq) \left(\frac{H}{p}\right)^8 \bar{A}_e'\!\cdot\!\bar{A}_e' \sin^2\cth \label{twop} \\
  &&\times x_p^4 x_q^4 \frac{\ln^2(2 \xi x_p)}{\ln(2 \xi x_p^e)} \frac{\ln^2(2 \xi x_q)}{\ln(2 \xi x_q^e)} \, , \nn\\
\nn\\
  \la \rho_\vp(x_p) \rho_\vq(x_q) \rho_\vk(x_k) \ra &=& - \frac{\delta^3(\vp+\vq+\vz)}{(2\pi)^3 a_{x_p}^4 a_{x_q}^4 a_{x_k}^4} \bar{A}_i'(x_p) \bar{A}_j'(x_q) \cep_+^i(\vp) \cep_+^j(\vq) \cep_-^k(\vp) \cep_-^k(\vq) \nn\\
  && \times \cA'(p,x_p) \cA'(q,x_q) \cA'(p,x_k) \cA'(q,x_k) + 2 \, \text{perm} \nn\\
  &=& 4pq\xi^4\cW^4 \frac{\delta^3(\vp+\vq+\vk)}{(2\pi)^3} \left(\frac{H^3}{pqk}\right)^4 \bar{A}_e'\!\star\!\bar{A}_e' \label{threep} \\
  && \times {x_p}^4 {x_q}^4 {x_k}^4 \frac{\ln^2(2 \xi x_p)}{\ln(2 \xi x_p^e)} \frac{\ln^2(2 \xi x_q)}{\ln(2 \xi x_q^e)} \ln(2 \xi_p x_k) \ln(2 \xi_q x_k) + 2 \, \text{perm} \, , \nn
\ea
where we defined $\xi_p \equiv \xi p/k$ and $\xi_q \equiv \xi q/k$, and we have used our solutions~(\ref{cA}); the permutations are obtained by looping over $x_p \rar x_q \rar x_k$.  Notice that in the case of $\bar{A}_i(x)$ we have used $\bar{A}_i(x) = \bar{A}_i(x_e) \ln(2 \xi x)/\ln(2 \xi x_e)$, as the background of infrared modes depends on time; the latter is drawn from an (approximately) Gaussian distribution with variance $\la \bar{A}_i'\bar{A}_i' \ra \sim 2a^4\rho_\text{cl}$ --- the classical energy density is taken from~(\ref{energy}) but integrating only for superhorizon modes at each given time.  Finally, we have used that $a(x) = k/Hx$ with the physical Hubble parameter defined through $aH=\cH$.  The $e$ subscript calls for a quantity to be evaluated at the end of inflation.  We have employed the shorthand notation $\bar{A}_e\!\star\!\bar{A}_e \equiv \bar{A}_i^e \, \bar{A}_j^e \, \cep_+^i(\vp) \cep_+^j(\vq) \cep_-^k(\vp) \cep_-^k(\vq)$.  In the same fashion, we can write the correlations among $\Pi$.

In the source term the time derivatives of the anisotropic stress and energy density appear; we can compute these by taking the synchronous limits of their two-point functions $x_1=x_2=x$ and looking at their square roots.  Therefore, we obtain $\la \rho^2 \ra \propto x^8 \ln^4(2 \xi x)$ modulo some $p/k$, $q/p$, and $k/q$ factors.  This means that, for example, $x\rho_{,x} \sim 4\rho - 2\rho/\ln(2 \xi x) \rar 4\rho$, for even if $\xi\gtrsim1$ we expect $\xi x \ll 1 \Rar \ln(2 \xi x) \gtrsim 1$.  However, notice that because of the transversality of the gauge field, the combination $(\rho+\Pi)$ does not contain leading order terms; within the approximation we employ, we can ignore it in computing the anisotropic spectrum and bispectrum --- it would be present in the case one were to look at the convolved spectrum, see below.

These expressions are to be inserted in the curvature spectrum and bispectrum, and integrated over time; we can discard the initial contribution (at horizon exit $x_h\approx1$) as it is small, even though the growth is only logarithmic.  Thence:
\bea
  &\int\!&\!\frac{\dd x_p}{x_p}\frac{\dd x_q}{x_q} x_p^4 x_q^4 \frac{\ln^2(2 \xi x_p)}{\ln(2 \xi x_q^e)} \frac{\ln^2(2 \xi x_q)}{\ln(2 \xi x_q^e)} \approx \frac{1}{4\times8} x^8 \frac{\ln^4(2 \xi x)}{\ln^2(2 \xi x_e)} = \frac{1}{32} x_e^8 \ln^2(2 \xi x_e) \, . \label{twoti} \\
\nn\\
  &\int\!&\!\frac{\dd x_p}{x_p}\frac{\dd x_q}{x_q}\frac{\dd x_k}{x_k} {x_p}^4 {x_q}^4 {x_k}^4 \frac{\ln^2(2 \xi x_p)}{\ln(2 \xi x_p^e)} \frac{\ln^2(2 \xi x_q)}{\ln(2 \xi x_q^e)} \ln(2 \xi_p x_k) \ln(2 \xi_q x_k) + 2 \, \text{perm} \nn \\
  && \approx \frac{1}{4\times8\times12} x^{12} \left(\frac{k}{q}\right)^4 \left(\frac{q}{p}\right)^8 \left(\frac{p}{k}\right)^{12} \frac{\ln^3(2 \xi_p x)}{\ln(2 \xi_p x_e)} \frac{\ln^3(2 \xi_q x)}{\ln(2 \xi_q x_e)} + 2 \, \text{perm} \nn \\
  && = \frac{1}{384} x_e^{12} \frac{p^4 q^4}{k^8} \ln^2(2 \xi_p x_e) \ln^2(2 \xi_q x_e) + 2 \, \text{perm} \, . \label{threeti}
\ea
The integration limits are $x_k = [1,kx_q/q]$, $x_q = [1,qx_p/p]$, and $x_p = [1,px/k]$, where in the end $x \rar x_e$.

All in all, the anisotropic two- and three-point functions now read
\bea
  \la \zeta_\vp \zeta_\vq \ra &\simeq& 2\pi^3 \left(\frac{1}{\epsilon\bar{\rho}}\right)^2 \xi^2\cW^2H^8 \frac{p}{\cH^8} \ln^2(2 \xi p/\cH) (\bar{A}_e')^2 \sin^2\cth \, \delta^3(\vp+\vq) \label{twozf} \, , \\
  \la \zeta_\vp \zeta_\vq \zeta_\vk \ra &\simeq& \frac{1}{3\times2^5 \pi^3} \left(\frac{1}{\epsilon\bar{\rho}}\right)^3 \xi^4\cW^4H^{12} \frac{pq}{\cH^{12}} \ln^2(2 \xi p/\cH) \ln^2(2 \xi q/\cH) \bar{A}_e'\!\star\!\bar{A}_e' \nn \\
  &&\times \delta^3(\vp+\vq+\vk) + 2 \, \text{perm} \label{threezf} \, .
\ea
The results quoted in the introduction are obtained by replacing
\[
  \bar{A}_e'\!\cdot\!\bar{A}_e' \sim 2a^4\rho \sim 16\xi^6\cW^2\cH^4\ln^2(2\xi) / \pi^2 \, ,
\]
and writing out the angles as $\bar{A}\!\star\!\bar{A} \equiv (\bar{A}\!\cdot\!\bar{A}) \, \mathfrak{A}(\cth_p,\cth_q,\cth')$ with $\cth_{(p,q,k)}$ the angles between the vector $\bar{A}$ and $(\vp,\vq,\vk)$, and $\cos\cth' = \hat{p}\!\cdot\!\hat{q}$.

\section{Phenomenology}\label{s:disc}

The power spectrum needs to be weighted against the isotropic part which comes from the free fluctuations of the inflaton field itself, and which evaluates to
\bea\label{p0}
  \la \zeta_\vp \zeta_\vq \ra_\cph \equiv 2\pi^2 \cP_0 \frac{\delta^3(\vp+\vq)}{p^3} = \frac{3H^4}{4\epsilon\bar{\rho}} \frac{\delta^3(\vp+\vq)}{p^3} \, ,
\ea
with amplitude $\cP_0 \approx 10^{-9}$, and from which we write out the total power spectrum
\bea\label{power}
  \cP(\vp)_\text{full} = \cP_0 \left[ 1 + \delta\cP(\vp) \right] \simeq \cP_0 \left[ 1 + \frac{1}{\cP_0} \frac{9\pi^3}{4\xi^4} \left( \frac{\rho}{\epsilon\bar{\rho}} \right)^2 \left(\frac{p}{\cH}\right)^4 \frac{\ln^2(2\xi p/\cH)}{\ln^2(2\xi)} \sin^2\cth \right] \, .
\ea
Similarly, the bispectrum can be weighted against the square of the spectrum similarly to the local non-Gaussianity template, albeit with a fully momenta-dependent shape:
\bea
  \la \zeta_\vp \zeta_\vq \zeta_\vk \ra \equiv \cB_{\bar{A}}(\vp,\vq,\vk) \, \delta^3(\vp+\vq+\vk) \equiv \frac{3}{10} f_\text{NL} \cP_0^2 \frac{p^3+q^3+k^3}{p^3q^3k^3} \, \delta^3(\vp+\vq+\vk) \, . \label{fnl}
\ea

To be concrete, we will examine two exemplifying cases.  First, we consider the scenario where the inflaton is the only field generating the dominant isotropic curvature perturbations; then we will consider the case where we allow for a low reheating temperature, which calls for the existence of a (large) number of curvaton fields.

\subsection{High scale N-flation}

If the inflaton is to provide for the isotropic part then we can use $\cP_0 = 3H^4/8\pi^2\epsilon\bar{\rho}$ from~(\ref{p0}) and obtain
\bea
  \delta\cP &=& 2^8\pi \cP_0 \left(\frac{p}{\cH}\right)^4 \ln^2(2\xi p/\cH) e^{4\pi\xi} \xi^6 \ln^2(2\xi) \sin^2\cth \, , \label{dpINF} \\
  f_\text{NL} &=& \frac{5\times2^5}{3^5\pi^2} \cP_0 \frac{k^3}{p^3+q^3+k^3} \frac{p^4q^4}{\cH^8} \ln^2(2\xi p/\cH) \ln^2(2\xi q/\cH) e^{6\pi\xi} \xi^7 \ln^2(2\xi) \mathfrak{A} + 2 \, \text{perm} \, . \label{fnlINF}
\ea
We are forced to fix a maximal value for $\xi$ in order for the slow-roll parameter $\epsilon$ to remain small, since it receives a contribution form the energy density of the gauge field.  This implies that, roughly, $\rho/\epsilon\bar{\rho}\ll1$ and, upon using~(\ref{energy}), we find approximately $\xi \lesssim 2.3$.

With this value, it is possible to check when the anisotropy rises to the order 1 level (or to any other desired): in this case this happens when $p/\cH \sim 10^{-3}$ at the end of inflation.  Having fixed the amplitude of $\cP_0$, and assuming instantaneous reheating, we are led to a scale of inflation which we can quantify through $H^2 = \pi^2 g_* T^4 / 90$ where $g_* = 106.75$ is the number of relativistic degrees of freedom which populate the Universe at the Big Bang.  This gives $T \approx 7.0\cdot10^{15}~\GeV$.  We then employ the relation $p/\cH \simeq 1.8\cdot10^{-8} p~\Mpc/(T/\GeV)$ to see that the anisotropy is confined to extremely small scales --- with these numbers, $p\approx10\text{cm}$, certainly not within observational reach.

The steepness of the anisotropic spectrum is due to the way the mechanism operates.  Inasmuch as most of the action takes place around horizon exit, and since all modes are almost adiabatically suppressed afterwards, like in vacuum, the most important contribution is entrusted to the high energy modes.  If the inflaton alone generates scalar perturbations, the scale of inflation is typically high, and the anisotropy peaks for very high momenta.  Similarly, the bispectrum~(\ref{fnlINF}) in this scenario has a coefficient of approximately $10^{-3}$ in front of the (unnormalised) shape; the latter is always much smaller than one except for the very last momenta to exit the horizon $p\sim q\sim k\sim \cH$, making the anisotropic bispectrum completely irrelevant.  The departures from vacuum in this case are not very strong.

Any application to magnetogenesis is similarly constrained.  If the $U(1)$ gauge field is identified with the Standard Model electromagnetic field (or some fraction of it), the spectrum of the magnetic field only at the end of inflation is given by
\bea\label{spectrum}
  \cP_\text{B}(p) = \frac{p^4}{8\pi^2a_e^4}\cW^2 = \frac{p^4}{16\pi^3a_e^4} \frac{e^{2\pi\xi}}{\xi} \, .
\ea
For a very rough estimate, we can simply assume that the field scales adiabatically, on large scales, until the present epoch.  This neglects all plasma phenomena such as cascades and inverse cascades, and diffusion.  Given that the field will be helical, for instance, there will be some inverse cascade effect, and power will be transferred from smaller scales to larger ones (see for a recent review~\cite{Durrer:2013pga}).  Also, on small scales diffusion will be in effect.  However, for the range we are dealing with (at around a Mpc), this estimation will suffice --- such scales only re-enter the Hubble horizon around or after recombination, when most of those plasma effects are not active any more.  Then today's power at a given wavelength is simply~(\ref{spectrum}) with $a_e\rar1$.  It is clear that, with the limiting value of $\xi\approx2.3$ there is no hope of generating any significant B field at large scales: the field strength on Mpc scales is below $10^{-47}~\nGauss$.

\subsection{Low scale N-flation}

We turn now to the opposite case, reversing the logic of the previous paragraph.  We could demand that the final B field is, for example, of order of 1 nGauss at a scale of about 1 Mpc.  The necessary amplification is attainable if $\xi\approx37$.  However, in this case, if we are to keep the ratio of magnetic (and electric) field to inflaton energy densities low, the spectrum of scalar curvature perturbations generated by the inflaton would be extremely small.  Thus, there needs be some other field, for example one or many curvatons, which feed $\cP_0$.  We are no longer allowed to use $\cP_0 = 3H^4/8\pi^2\epsilon\bar{\rho}$.

Assuming this to be the case, one needs to practically ensure that backreaction does not set for a sufficient period of inflation, see for instance~\cite{Demozzi:2009fu,Kanno:2009ei,Urban:2011bu}.  Since $\xi$ is fixed, one can play with the reheating temperature.  The ratio $\rho/\epsilon\bar{\rho}$ remains small if now the reheating temperature is small, that is, if $T\lesssim1~\GeV$, which is extremely low.

Looking again at the anisotropy, writing explicitly $H^2 = \pi^2 g_* T^4 / 90$, and knowing that $p/\cH \simeq 1.8\cdot10^{-8} p~\Mpc/(T/\GeV)$ in the usual relations
\bea
  \delta\cP &=& \frac{1}{\cP_0} \frac{9\pi^3}{4\xi^4} \left(\frac{\rho}{\epsilon\bar{\rho}}\right)^2 \left(\frac{p}{\cH}\right)^4 \frac{\ln^2(2\xi p/\cH)}{\ln^2(2\xi)} \sin^2\cth \, , \label{dpCURV} \\
  f_\text{NL} &=& \frac{1}{\cP_0^2} \frac{k^3}{p^3+q^3+k^3} \frac{p^4q^4}{\cH^8} \frac{5\pi}{2^{10} 9 \xi^8} \left(\frac{\rho}{\epsilon\bar{\rho}}\right)^3 \frac{\ln^2(2 \xi p/\cH) \ln^2(2 \xi q/\cH)}{\xi^8 \ln^4(2\xi)} \mathfrak{A} + 2 \, \text{perm} \, , \label{fnlCURV}
\ea
we realise that the anisotropy is maximal at the maximum temperature we can afford, and again at the highest momentum processed by inflation.  However, the actual value of $\delta\cP$ blows up very quickly because of the large $\xi$ bringing about a great exponential enhancement.  In numbers, we reach order one anisotropy at $p\approx6\times10^{-3}/\Mpc$, which is not acceptable, especially knowing that it grows with the quartic power of momentum itself on smaller scales.  The only way to reduce this is to further lower the reheating temperature, but even with the lowest allowed value of about 1 MeV the anisotropy on large scales remains large: $\delta\cP\approx1$ for $p\approx2/\Mpc$.

Unfortunately, were we able to prevent a large anisotropy from forming, this still would not be a viable situation for magnetogenesis.  The steep spectral slope of~(\ref{spectrum}) implies that while at around 1 Mpc the strength of the magnetic field is compatible with observations, certainly this would not be the case at smaller scales, where it blows up quartically --- it is not possible to have fields much stronger than about a nGauss even several orders of magnitude below the Mpc scale~\cite{Durrer:2013pga}.

One could ask which is the limiting value for $\xi$ at a given temperature for which the anisotropy is of order 1 --- this irrespectively on whether the gauge field is the electromagnetic field.  With the anisotropy strongly peaking at small scales, one could for instance demand that it does not become important until beyond some ``safe'' ultraviolet cutoff.  For example, let us pick $p\sim10^4/\Mpc$.  Choosing again 1 GeV as our reheating temperature, from~(\ref{dpCURV}) we find that $\xi\approx32$.  Identifying the gauge field with electromagnetism, the corresponding magnetic field strength would be small at that scale, but would lie in the interesting observable range: $10^{-6}~\nGauss$ at 1 Mpc.  Nevertheless, the magnetic power on, say, pc scales, would still be too large.  Notice finally that in this case the ratio $\rho/\bar{\rho}$ is still very small, only $10^{-13}$.

To conclude: as soon as the temperature of inflation is lowered, in principle larger values of the coupling $\xi$ are allowed from a backreaction point of view.  However, the anisotropic spectrum is strongly enhanced, due to the exponential factor $\exp(2\pi\xi)$, and presents a steep dependence on $p/\cH$: the anisotropy, and the corresponding $f_\text{NL}$ also tend to blow up rapidly.  Magnetogenesis is strongly inhibited by this argument; moreover, the steepness of the $U(1)$ spectrum favours power on smaller scales, making it very hard to have appreciable fields on large scales and obey all constraints on even slightly smaller scales.

\subsection{Remarks}

An interesting comparison can be made by looking at what the curvature power spectrum would have looked like if we ignored the existence of the anisotropic background, which amounts to neglecting the $\bar{A}$ in the expansion for the vector potential~(\ref{cA}).  The outcome is
\[
  \la \zeta_\vp \zeta_\vq \ra_\text{iso} \sim \frac{1}{\xi^3} \left(\frac{\rho}{\epsilon\bar{\rho}}\right)^2 \left(\frac{p}{\cH}\right)^3 \, \frac{\delta^3(\vp+\vq)}{p^3} \, .
\]
This spectrum is isotropic, but is strongly tilted.  Of course in principle it is conceivable that a different coupling function $f(\cph)$ can return an almost scale-invariant spectrum, but we are interested here in noticing the difference with~(\ref{twozEND}).  The amplitudes are not dramatically different; there is a small extra parametric suppression in~(\ref{twozEND}) with the coupling $\xi$.  The interesting feature is in the spectral shape: the unphysical isotropic spectrum has a different slope (in this case, shallower) than the physical anisotropic one.  We expect this to be the case in general scenarios --- this indeed happens in~\cite{Bartolo:2012sd,Urban:2013aka}.  Thus, the isotropic spectrum not only misses the angular dependence of the physical result, but also misleads in terms of the general structure with momentum.

Incidentally, an inflationary model in which friction, responsible for the slow rolling of the inflaton, comes from the gauge field dissipation term (rather than the expansion of the Universe versus the flatness of the potential) as in~\cite{Anber:2009ua,Anber:2012du}, is destined to possess an inherently anisotropic curvature power spectrum.  Notice that in~\cite{Anber:2009ua,Anber:2012du} the limit for which $2\sqrt{2 \xi x} \gg 1$ is investigated, so that only the first lines of~(\ref{g0}), (\ref{f0}), (\ref{dg0}), and~(\ref{df0}) make sense (since still $x\ll2\xi$), so we can not directly lend our results to that case.  One further complication comes from the fact that the slow-roll approximations we employ in deriving~(\ref{zetadyn}) do not all apply.

\subsection{Conclusion}

We have examined the effects of the axial coupling between a (number of) pseudoscalar inflaton and a $U(1)$ vector field, looking for its effect on the curvature power spectrum and bispectrum.  The sourced curvature turns out to be anisotropic, and strongly tilted towards high energy modes.  We have identified the origin of such behaviour in the evolution of the coupling function --- only active around horizon crossing --- and discussed the detectability of the anisotropy.  As it turns out, even for energy densities of the gauge field approaching that of the isotropic background, the anisotropy seems to be too small to be detected; this is particularly so since if anywhere, the anisotropy would strongly prefer small scales, which are out of reach of current experiments.  This is certainly the case for N-flation alone, where the first order isotropic curvature perturbation is due to the multitude of pseudoscalar themselves; the coupling is forced to remain small there in order to not spoil the background, and the departures from vacuum are small.

If the scale of inflation is low and the isotropic curvature perturbations are generated by some other fields, then more interesting results can be obtained.  This applies to both the spectrum and the bispectrum.  In this case it would be in principle conceivable that the gauge field is nothing else than the familiar Maxwell field, and applications to large scale magnetogenesis would be possible; however, the magnetic field spectrum is very blue, and large power on large scales is only attained at the cost of much larger power on smaller scales, which does not seem compatible with observations.  Moreover, in this case the anisotropy is also boosted due to its spectral dependence, and one needs to adjust the parameters of the model in order to keep it small.  In general, the smaller the anisotropy at a given scale, and the smaller the generated magnetic field is.  Magnetogenesis with Mpc sized fields in the pico Gauss range appears to be compatible with a low degree of statistical anisotropy at large, observable, scales, but not with the strength of the magnetic field itself on even mildly smaller scales.

Still, with a generic $U(1)$ field there is a region in the parameters space where one there phenomenology can be quite rich.  This however demands a mechanism which keeps the number of fields high --- so as to have a large coupling ---, and the scale of inflation substantially low --- so as to keep the anisotropy under control.  Since in this case some curvaton mechanism is needed, it remains to be seen whether there are other interesting effects.  In particular it would be natural to expect some kind of interaction between the curvatons themselves and the gauge field.  We leave this issue for future investigation.

\section*{Acknowledgements}
FU wishes to thank Lorenzo Sorbo for useful correspondence.  He is supported by IISN project No.~4.4502.13 and by Belgian Science Policy under IAP VII/37.

\bibliography{ax}
\bibliographystyle{JHEP}

\end{document}